\documentstyle [12pt] {article}
\begin{document}

\pagestyle{empty}
\rightline{\vbox{\halign{&#\hfil\cr
	&NUHEP-TH-94-21 \cr
	&September 1994 \cr
	&hep-ph/9410218 \cr}}}
\bigskip
\bigskip
\bigskip
{\Large\bf
	\centerline{Asymptotic Behavior of the}
	\centerline{Correlator for Polyakov Loops}}
\bigskip
\normalsize

\centerline{Eric Braaten and Agustin Nieto}
\centerline{\sl Department of Physics and Astronomy, Northwestern University,
    Evanston, IL 60208}
\bigskip

\begin{abstract}
The asymptotic behavior of the correlator for Polyakov loop operators
separated by a large distance $R$ is determined for high temperature QCD.
It is dominated by nonperturbative effects
related to the exchange of magnetostatic gluons.
To analyze the asymptotic behavior, the problem is formulated in terms
of the effective field theory of QCD in 3 space dimensions.
The Polyakov loop operator
is expanded in terms of local gauge-invariant operators constructed out
of the magnetostatic gauge field, with coefficients that
can be calculated using resummed perturbation theory.
The asymptotic behavior of the correlator is $\exp(-MR)/R$,
where $M$ is the mass of the lowest-lying glueball in $(2+1)$-dimensional
QCD.  This result implies that existing lattice calculations
of the Polyakov loop correlator at the highest temperatures available
do not probe the true asymptotic region in $R$.
\end{abstract}

\vfill\eject\pagestyle{plain}\setcounter{page}{1}

One of the basic characteristics of a plasma is the
screening of electric fields.  The field created by a
static charge falls off exponentially beyond the screening radius,
whose inverse is called the Debye mass $m_D$.
In a quark-gluon plasma at high temperature $T$,
chromoelectric fields are believed to be screened in a similar way.
However it has proven to be difficult to give a precise definition
to the Debye mass in perturbation theory.
At leading order in the coupling constant $g$, $m_D$ is proportional to
$g T$.  The next-to-leading order correction to $m_D$ has been calculated
using a resummed perturbation theory in which gluon propagator
corrections of order $g^2T^2$ are summed up to all orders \cite{rebhan,bn}.
The correction is gauge-invariant but infrared-divergent, indicating a
sensitivity to nonperturbative effects involving the scale $g^2 T$.

Debye screening can also be studied nonperturbatively using lattice
simulations \cite{attig,gao,kark}.
One of the simplest probes of Debye screening
is the correlator of two Polyakov loop operators as a function of their
separation $R$.  At lowest order in resummed perturbation theory,
the behavior of the correlator is predicted to be
$\exp (- 2 m_D R)/R^2$.  By fitting the measured correlator to this
form, one can extract a value for $m_D$.  However the resummed
perturbation expansion is known to break down at higher orders,
due to contributions that involve the exchange of magnetostatic gluons.
These contributions, which are inherently nonperturbative, can be
expected to dominate at distances $R$ much greater than $1/m_D$.
This raises questions about the utility of determining $m_D$ by
fitting the correlator to a leading order expression.

In this paper, we determine the true asymptotic behavior of the
Polyakov loop correlator at large $R$.  We use effective field theory
methods to express the Polyakov loop operator in terms of operators
in 3-dimensional QCD.  The asymptotic behavior of the Polyakov loop
correlator is then given by a
simple correlator in this effective theory.  It has the form
$\exp(- m_H R)/R$, where $m_H$ is the mass of the lowest
glueball state in (2+1)-dimensional QCD.  The coefficient
of the exponential is of order $g^{12}$.  Our result implies that
the asymptotic behavior of the Polyakov loop correlator is dominated
by magnetostatic effects which have little to do with Debye screening.

We wish to study the correlator of Polyakov loop operators in thermal QCD
in 4 space-time dimensions with temperature $T$.
The fundamental fields are the
gauge field $A_\mu({\bf x},\tau)$, which takes values in the
$SU(N_c)$ algebra, and the quark fields $\psi^i({\bf x}, \tau)$,
whose indices range over $N_c$ colors and $N_f$ flavors.
The gluon fields satisfy periodic boundary conditions in the
Euclidean time $\tau$ with period $\beta = 1/T$,
while the quark fields obey antiperiodic boundary conditions.
The Polyakov loop operator is given by the trace of a
path-ordered exponential:
\begin{equation}
L({\bf x}) \;=\; {1 \over N_c} {\rm tr} \;
{\cal P} \exp \left( - i g \int_0^\beta d \tau A_0({\bf x},\tau) \right) .
\label{PLO}
\end{equation}
The connected part of the correlator of two Polyakov loop operators
separated by a distance $R$ is
\begin{equation}
C(R) \;=\;
\langle L^\dagger({\bf 0}) \; L({\bf R}) \rangle
	\;-\; \langle L({\bf 0}) \rangle^2 .
\label{CPL}
\end{equation}

The Polyakov loop operator (\ref{PLO}) creates two or more electric gluons
($A_0$ fields).  The diagram for the Polyakov loop correlator (\ref{CPL})
which is leading order in a naive expansion in powers of $g$ is the
1--loop diagram in which two electrostatic gluons are exchanged.
It gives the asymptotic behavior $C(R) \sim 1/R^2$.
This is not the correct asymptotic behavior for two reasons.  First,
thermal loop corrections generate a Debye screening mass for
electrostatic gluons, so that the potential due to the exchange of an
electrostatic gluon actually falls off exponentially, rather than
like $1/R$ as in naive perturbation theory.
Secondly, the true asymptotic behavior
at sufficiently large temperature actually comes from higher order diagrams
that involve the exchange of magnetostatic gluons.
Our problem is to determine this asymptotic behavior.

An elegant way to solve this problem is to construct a sequence of two
effective field theories which reproduce static correlation functions
at successively longer distances.  Thermal QCD can be used to
calculate the static correlator (\ref{CPL}) for any separation $R$.
Ordinary perturbation theory in $g^2$ is accurate only for $R$
of order $1/T$ or smaller, but the correlator
can be calculated for larger $R$ by using nonperturbative
methods such as lattice gauge theory simulations.
The first of the two effective field theories is constructed so that it
reproduces static correlators at distances $R$ of order $1/(gT)$ or larger.
In this effective theory, perturbation theory in $g$ can be used to calculate
correlators at distances of order $1/(gT)$, but lattice simulations are
required at larger $R$.  The second effective field theory
is constructed so that it reproduces correlators at distances of
order $1/(g^2 T)$ or larger.
This field theory is completely nonperturbative and the correlators must
be calculated by lattice simulations.  Nevertheless, we can use
this field theory to determine unambiguously the asymptotic behavior
of the Polyakov loop correlator.

The first effective field theory, which we call electrostatic QCD (EQCD),
is a 3-dimensional Euclidean field theory that contains
the electrostatic gluon field $A_0({\bf x})$ and the magnetostatic
gluon field $A_i({\bf x})$.  Up to a normalization, they can be identified
with the zero-frequency modes of the gluon field $A_\mu({\bf x}, \tau)$
for thermal QCD in a static gauge \cite{nadkarni1}.
The action for the effective field theory is
\begin{equation}
S_{\rm EQCD} \;=\; \int d^3x \bigg\{
{1 \over 2} {\rm tr} (G_{ij} G_{ij})
	\;+\; {\rm tr} (D_i A_0 D_i A_0)
	\;+\; m_{\rm el}^2 \; {\rm tr} (A_0 A_0)
	\;+\; \delta {\cal L}_{\rm EQCD} \bigg\} ,
\label{EQCD}
\end{equation}
where $G_{ij} = \partial_i A_j - \partial_j A_i + ig_3 [A_i,A_j]$ is the
magnetostatic field strength,
$D_i A_0 = \partial_i A_0 + i g_3 [A_i,A_0]$,
and $g_3$ is the coupling constant of the
3-dimensional gauge theory.  The action for this effective
field theory is invariant under static $SU(N_c)$ gauge transformations.
If the fields $A_0$ and $A_i$ are assigned
dimension $1/2$, then the operators shown explicitly in (\ref{EQCD})
have dimensions 3, 3, and 1.   The term $\delta {\cal L}_{\rm EQCD}$
in (\ref{EQCD}) includes all other local gauge-invariant operators
of dimension 2 and higher that can be constructed out of $A_0$ and $A_i$.
The effective theory EQCD is completely equivalent to thermal QCD
at distance scales of order $1/(gT)$ or larger.  The gauge coupling constant
$g_3$, the mass parameter $m_{\rm el}^2$, and the
parameters in $\delta {\cal L}_{\rm EQCD}$ can be tuned as functions of $T$
so that correlators of gauge invariant operators in EQCD
agree with the corresponding static correlators in thermal QCD to any
desired accuracy for $R \gg 1/T$ \cite{lepage}.
By matching the two theories at tree level,
we find that the gauge coupling constant is $g_3 = g(T) \sqrt{T}$,
where $g(T)$ is the running coupling constant at the momentum scale $T$.
The mass parameter $m_{\rm el}^2$ in (\ref{EQCD}) is the contribution
to the square of the Debye screening mass from short distances
of order $1/T$.  At leading order in $g$, it is
\begin{equation}
m_{\rm el}^2 \;=\;  {2 N_c + N_f \over 6} g^2(T) T^2 .
\label{mel}
\end{equation}
The coefficients of some of the higher dimension operators in
$\delta {\cal L}_{\rm EQCD}$
were recently calculated to leading order by Chapman \cite{chapman}.
Since the parameters in EQCD only take into account the effects of
the momentum scale $T$, they can be calculated as perturbation series in
$g^2(T)$.  For example, the next-to-leading order correction
to $m_{\rm el}^2$ is of order $g^4(T)$.
The Debye screening mass $m_D$ defined by the location of the pole
in the gluon propagator is also given at leading order by (\ref{mel}),
but $m_D^2$ has corrections of order $g^3$ that arise from
the momentum scale $gT$ \cite{rebhan,bn}.

The effective field theory EQCD was used by Nadkarni to study the Polyakov
loop correlator beyond leading order in the coupling constant \cite{nadkarni2}.
In EQCD, the Polyakov loop operator is given by a simple exponential:
\begin{equation}
L({\bf x}) \;=\; {1 \over N_c}
{\rm tr} \; \exp \left( - i g A_0({\bf x}) / \sqrt{T} \right).
\end{equation}
The 1--loop diagram involving the exchange of two
electrostatic gluons gives a correlator that falls exponentially due
to electric screening:
\begin{equation}
C(R) \;\longrightarrow\;
{(N_c^2 - 1) g^4 \over 8 N_c^2 T^2}
	\left( {e^{- m_{\rm el} R} \over 4 \pi R} \right)^2 .
\label{CEas}
\end{equation}
The corrections to this correlator of order $g^6$ were
calculated by Nadkarni \cite{nadkarni2}.  They have the asymptotic
behavior $e^{-2 m_{\rm el} R} \log(R)/R$.  In a recent reexamination of this
calculation \cite{bn},  it has been shown that one can extract from it
the correction of order $g^3$ to the square of the Debye mass that was first
obtained by Rebhan from the pole in the gluon propagator \cite{rebhan}.

As pointed out by Nadkarni \cite{nadkarni2},  the asymptotic behavior
of the Polyakov loop correlator comes not from the 1--loop diagram
in which two electrostatic gluons are exchanged,
but instead from higher--loop diagrams that involve the exchange
of magnetostatic gluons.  The simplest such diagrams are
the 3--loop diagrams in Fig.~\ref{fig1}.
In perturbation theory, magnetostatic gluons remain massless in
EQCD, and the diagrams in Fig.~\ref{fig1} give a contribution to the
correlator that falls like $1/R^6$.  However this is not the correct
asymptotic behavior, since nonperturbative effects become important
at a distance $R$ of order $1/(g^2T)$.

In order to determine the true asymptotic behavior of the Polyakov
loop correlator, it is useful to construct a second effective field
that reproduces static correlators at distances  $R$ of order $1/(g^2T)$
or larger.  This effective theory, which we call magnetostatic QCD (MQCD),
is a pure SU(3) gauge theory in 3 space dimensions.
The only fields are the magnetostatic gluon fields $A_i({\bf x})$.
The action is
\begin{equation}
S_{\rm MQCD} \;=\; \int d^3x \left\{
{1 \over 2} {\rm tr} (G_{ij} G_{ij})
	\;+\; \delta {\cal L}_{\rm MQCD} \right\} ,
\label{MQCD}
\end{equation}
where $\delta {\cal L}_{\rm MQCD}$ includes all local
gauge-invariant operators that can be constructed out of $A_i$.
The gauge coupling constant of MQCD and the
parameters in $\delta {\cal L}_{\rm MQCD}$ can be tuned so that
MQCD is completely equivalent to EQCD, and therefore to thermal QCD,
at distances of order $1/(g^2T)$ or larger.
If $g(T)$ is sufficiently small, the parameters of MQCD can be obtained
by perturbative calculations in EQCD.  The expansion parameter is
$g_3^2/m_{\rm el}$,  which is of order $g(T)$.
The gauge theory MQCD itself is inherently nonperturbative.
Any perturbative expansion in powers of $g_3$ is hopelessly plagued
with infrared divergences.  Thus the correlation functions in
this theory must be calculated nonperturbatively using
lattice simulations.

The static correlators of gauge-invariant operators in EQCD
are reproduced at distances $R \gg 1/m_{\rm el}$ by the corresponding
operators in MQCD.
In EQCD, the Polyakov loop operator creates electrostatic gluons
only.  It couples to magnetostatic gluons through loop diagrams involving
electrostatic gluons, such as the 1--loop diagrams in Fig.~\ref{fig2}.
Because of screening, electrostatic gluons can only propagate over distances
of order $1/m_{\rm el}$.  Thus for magnetostatic gluons with wavelengths
much greater than $1/m_{\rm el}$, the Polyakov loop operator behaves like a
point-like operator that creates magnetostatic gluons.  It can
therefore be expanded out in terms of local gauge-invariant operators
constructed out of the field $A_i$:
\begin{equation}
L({\bf x}) \;=\;
\lambda_1(g) \; 1
\;+\; {\lambda_{G^2}(g) \over m_{\rm el}^3} \; G^2({\bf x}) \;+\; \ldots,
\label{PLope}
\end{equation}
where $G^2 \equiv {\rm tr} (G_{ij} G_{ij})$ and
$\ldots$ represents operators of dimension 5 or larger, such as
$G^3  \equiv g_3 {\rm tr} (G_{ij}G_{jk}G_{ki})$.

Like the parameters in the effective action (\ref{MQCD}),
the coefficients in the operator expansion (\ref{PLope}) are computable
in terms of the parameters of the EQCD action using perturbation theory in $g$.
Both EQCD and MQCD reproduce the nonperturbative dynamics
of thermal QCD at distances much greater than $1/m_{\rm el}$.
Their perturbation expansions also give equivalent
(although incorrect) descriptions of the long-distance dynamics.
Since the coefficients in the operator expansion (\ref{PLope})
are insensitive to the long-distance dynamics,
the equivalence between perturbative EQCD and perturbative
MQCD can be exploited as a device to compute the coefficients.

We proceed to calculate the coefficient $\lambda_{G^2}$ in (\ref{PLope})
to leading order in $g$.  The simplest quantity that can be used to
calculate $\lambda_{G^2}$ is the coupling of the Polyakov loop operator
to two long-wavelength magnetostatic gluons, which we denote by
$\langle 0 | L({\bf 0}) | gg \rangle$.  We take the gluons to have
momenta ${\bf k}_1$ and ${\bf k}_2$, vector indices $i$ and $j$,
and color indices $a$ and $b$.  In perturbative MQCD, we
can read off the coupling of the operator $L({\bf 0})$
to the two gluons directly from the expression (\ref{PLope}) for
the Polyakov loop operator:
\begin{equation}
\langle 0| L({\bf 0}) | gg \rangle
\;=\; {2 \lambda_{G^2} \over m_{\rm el}^3} \; \delta^{ab}
\left(- {\bf k}_1 \cdot {\bf k}_2 \delta^{ij} + k_2^i k_1^j \right).
\label{Omgg}
\end{equation}
In perturbative EQCD, the coupling is given by the sum of
the 1--loop diagrams in Fig.~\ref{fig2}:
\begin{eqnarray}
\langle 0 | L({\bf 0}) | gg \rangle \;=\;
{g^4 \over 2} \delta^{ab} \int {d^3p \over (2 \pi)^3}
	{1 \over {\bf p}^2 + m_{\rm el}^2}
	{1 \over ({\bf p} + {\bf k}_1 + {\bf k}_2)^2 + m_{\rm el}^2}
\nonumber \\
\left( \delta^{ij} + { (2 {\bf p} + {\bf k}_1)^i
		(2 {\bf p} + 2 {\bf k}_1 + {\bf k}_2)^j
		\over ({\bf p} + {\bf k}_1)^2 + m_{\rm el}^2} \right) .
\label{OmggE}
\end{eqnarray}
Expanding the integrand out to second order in ${\bf k}_1$ and ${\bf k}_2$
and evaluating the loop integrals, we find that (\ref{OmggE}) reduces to
\begin{equation}
\langle 0 | L({\bf 0}) | gg \rangle \;=\;
{g^4 \over 192 \pi m_{\rm el}^3} \delta^{ab}
\left(- {\bf k}_1 \cdot {\bf k}_2 \delta^{ij} + k_2^i k_1^j \right).
\label{Omggk}
\end{equation}
Comparing (\ref{Omgg}) and (\ref{Omggk}), we can read off the coefficient
$\lambda_{G^2}$:
\begin{equation}
\lambda_{G^2} \;=\; {g^4 \over 384 \pi}.
\label{C3}
\end{equation}

Having determined the coefficient $\lambda_{G^2}$ in the operator expansion
(\ref{PLope}) to leading order in $g$,  we can now express the
Polyakov loop correlator (\ref{CPL})
in terms of correlators in MQCD:
\begin{equation}
C(R) \;=\; \left( {\lambda_{G^2} \over m_{\rm el}^3} \right)^2
\langle G^2({\bf 0}) G^2({\bf R}) \rangle \;+\; \ldots \; .
\label{CGG}
\end{equation}
The $\ldots$ in (\ref{CGG}) represents the contributions of higher
dimension operators in the operator expansion (\ref{PLope}),
such as $G^3({\bf x})$.

The asymptotic behavior of a correlator in MQCD, such as the one
in (\ref{CGG}), is related to the spectrum of QCD in (2+1) space-time
dimensions.  This is a confining gauge theory with a dynamically generated
mass gap $M_H$ between the vacuum and the state of next lowest energy.
The single particle states in the spectrum are bound states of
gluons (glueballs), and $M_H$ is the mass of the lightest glueball.
Assuming that the lowest glueball $H$ is a scalar particle,
the Fourier transform of the correlator
$\langle G^2({\bf 0}) G^2({\bf R}) \rangle$ has a pole at $k^2 = - M_H^2$,
as well as poles and branch cuts that are farther from the real $k$ axis.
The asymptotic behavior in $R$ is dominated by the pole at $k^2 = - M_H^2$.
Denoting the residue of the pole by
$|\langle 0 | G^2({\bf 0}) | H \rangle|^2$, the asymptotic behavior is
\begin{equation}
C(R) \;\longrightarrow\; \left( {\lambda_{G^2} \over m_{\rm el}^3} \right)^2
|\langle 0 | G^2({\bf 0}) | H \rangle |^2  {e^{- M_H R}  \over 4 \pi R}.
\label{CMas}
\end{equation}
Both the mass $M_H$ and the coupling strength
$\langle 0 | G^2({\bf 0}) | H \rangle$ can be calculated using lattice
simulations of MQCD.  By dimensional analysis,
$M_H$ is proportional to $g^2 T$, while $\langle 0 | G^2({\bf 0}) | H \rangle$
is proportional to $(g^2T)^{5/2}$.  The overall coefficient of the exponential
in (\ref{CMas}) is therefore proportional  to $g^{12}$.

It is interesting to compare the asymptotic behavior (\ref{CMas})
with the result one would obtain at leading order in perturbation theory
in MQCD.  The leading order diagram is the 1--loop diagram in which
two magnetostatic gluons are exchanged.
The possibility of a dynamically-generated magnetic screening mass
$m_{\rm mag}$ can be taken into account by replacing the propagators
$1/{\bf p}^2$ of the gluons by $1/({\bf p}^2 + m_{\rm mag}^2)$.
The resulting expression for the correlator is
\begin{equation}
C(R) \;\approx\; 2 (N_c^2 -1)
\left( {\lambda_{G^2} \over m_{\rm el}^3} \right)^2
\left( {e^{- m_{\rm mag} R} \over 4 \pi R^3} \right)^2
\left[ 6 + 12 m_{\rm mag} R + 10 (m_{\rm mag} R)^2 + 4 (m_{\rm mag} R)^3
\right]
\end{equation}
The perturbative result, which is obtained by setting $m_{\rm mag} = 0$,
falls like $1/R^6$.
In the presence of magnetic screening, the asymptotic behavior is
$e^{-2 m_{\rm mag} R}/R^3$.  Thus this model does not give the same
asymptotic behavior as (\ref{CMas}), even if we make the identification
$m_H = 2 m_{\rm mag}$.

The correlator of Polyakov loops has been calculated nonperturbatively
using lattice simulations \cite{attig,gao,kark}.
Recent studies have found that, at temperatures well above the
quark--gluon plasma phase transition, the behavior of the correlator
at large $R$ is consistent with the form $e^{- \mu R}/R^n$
with $n$ in the range $1 < n < 2$.  At the highest temperatures
available, the preferred value of $n$ is close to 2, as
predicted by the leading electrostatic contribution (\ref{CEas}).
At lower temperatures, the preferred value of $n$ is closer
to 1, consistent with the true asymptotic result (\ref{CMas}).
These numerical results have a simple interpretation.
The electrostatic contribution (\ref{CEas})
falls off like $e^{-2 m_{\rm el} R}/(T R)^2$, with a coefficient
proportional to $g^4$.
The magnetostatic contribution (\ref{CMas}) falls off more slowly
like $e^{-M_H R}/(T R)$, but its coefficient
is proportional to $g^{12}$.  At the highest temperatures
that have been probed by lattice simulations, the running coupling constant
$g(T)$ is quite small.  Because it has a very small coefficient,
the magnetostatic contribution will probably not
dominate over the electrostatic contribution until $R$ is much larger
than the size of present lattices.  Thus the measured correlator behaves like
$e^{- \mu R}/R^2$, and the mass $\mu$ extracted by fitting the correlator
to this form can be interpreted as twice the Debye screening mass.
At lower temperatures, the magnetostatic contribution
is not so strongly suppressed and the true asymptotic behavior $e^{-\mu R}/R$
is probably observed on the lattice.  If this interpretation of
the numerical simulations is correct, then the mass $\mu$ extracted from
fitting the correlator to the form $e^{- \mu R}/R$ has nothing to do with
Debye screening, but instead is related to magnetostatic effects.
This interpretation could be verified by calculating $m_H$
and $\langle0|G^2({\bf 0})|H\rangle$ using a lattice simulation of MQCD.
One could then use (\ref{CMas})
to predict quantitatively how large $R$ must be in order to reach the truly
asymptotic region of the Polyakov loop correlator.

This work was supported in part by the U.S. Department of Energy,
Division of High Energy Physics, under Grant DE-FG02-91-ER40684,
and by the Ministerio de Educaci\'on y Ciencia of Spain.  We are grateful
to A. Rebhan for pointing out an error in a draft of this paper
which significantly altered the conclusions.

\vfill\eject

\newpage
\begin{center}\section*{Figure Captions}\end{center}

\begin{enumerate}

\item
\label{fig1}
Three--loop diagrams in EQCD that contribute to the correlator of
two Polyakov loop operators.  The solid lines are electrostatic gluons
and the wavy lines are magnetostatic gluons.

\item
\label{fig2}
One--loop diagrams in EQCD that couple a Polyakov
loop operator to two magnetostatic gluons.

\end{enumerate}

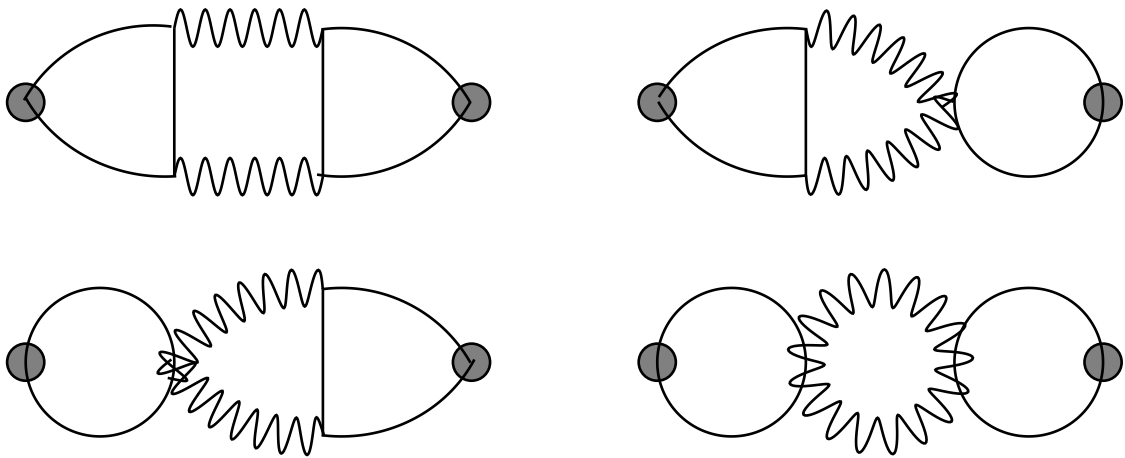
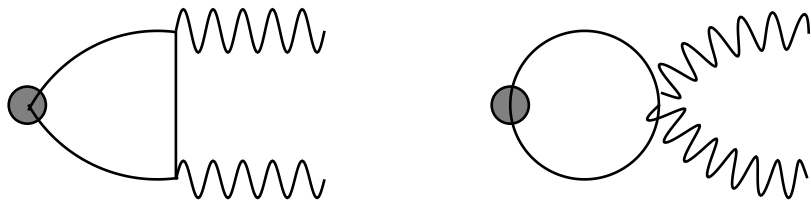
\end{document}